\documentclass[useAMS,usegraphicx]{mn2e}
\usepackage{rotate}
\usepackage{times}
\newif\ifAMStwofonts
\AMStwofontstrue

\def\gs{\mathrel{\hbox{\rlap{\hbox{\lower4pt\hbox{$\sim$}}}\hbox{$>$}}}}
\def\ls{\mathrel{\hbox{\rlap{\hbox{\lower4pt\hbox{$\sim$}}}\hbox{$<$}}}}

\def\einstein{{\it Einstein}}

\def\chandra{{\it Chandra}}

\def\hst{{\it HST}}
\def\rosat{{\it ROSAT}}

\def\heao{{\it HEAO-1} A2}
\def\asca{{\it ASCA}}

\def\xmm{{\it XMM-Newton}}

\def\lobster{{\it Lobster}}

\def\et{{et al.\ }}
\def\ic{{IC~3599}}
\def\wpvs{{WPVS~007}}
\def\rxj{{RX~J1624.9+7554}}
\def\rj{{RX~J1242.6-1119}}
\def\ngc{{NGC~5905}}
\def\hii{{H~\textsc{ii}}}

\def\arcsec{{\hbox{$^{\prime\prime}$}}}

\def\km{{\rm\thinspace km}}

\def\Mpc{{\rm\thinspace Mpc}}

\def\ps{{\rm\thinspace s^{-1}}}

\def\kmps{\hbox{$\km\ps\,$}}
\def\kmpspMpc{\hbox{$\kmps\Mpc^{-1}\,$}}

\title[\chandra\ observations of five X-ray transient galactic nuclei ]
      {\chandra\ observations of five X-ray transient galactic nuclei}

\author[Vaughan, Edelson \& Warwick]
       {S. Vaughan,$^{1,2}$
        R. Edelson$^{3}$ and
        R. S. Warwick$^{2}$ \\
$^{1}$Institute of Astronomy; Madingley Road; Cambridge, CB3 0HA\\
$^{2}$X-Ray Astronomy Group; University of Leicester; Leicester, LE1
7RH\\
$^{3}$Astronomy Department; University of California; Los Angeles, CA
90095-1562; USA }

\date{Accepted: 14/1/2004; Submitted: 23/12/2003; in original form: 12/11/2003}
\pagerange{\pageref{firstpage}--\pageref{lastpage}}

\pubyear{2003}
\begin{document}
\maketitle
\label{firstpage}

\begin{abstract}
We report on exploratory \chandra\ observations of five galactic
nuclei that were found to be X-ray bright during the \rosat\ all-sky
survey (with $L_{\rm X} \gs 10^{43}$ erg s$^{-1}$) but subsequently
exhibited a dramatic decline in X-ray luminosity.   Very little is
known about the post-outburst X-ray properties of these enigmatic
sources.   In all five cases \chandra\ detects an X-ray source
positionally coincident with the nucleus of the  host galaxy.   The
spectrum of the brightest source (\ic) appears consistent with a steep
power-law ($ \Gamma \sim 3.6 $).   The other sources have too few
counts to extract individual, well-determined spectra, but their X-ray
spectra appear flatter ($ \Gamma \sim 2 $) on average.   
The \chandra\ fluxes are  $\sim 10^2-10^3$ fainter than was observed
during the outburst (up to $12$ years previously). That all
post-outburst X-ray observations showed similarly low
X-ray luminosities is consistent with these sources having `switched' to a
persistent low-luminosity state. Unfortunately the relative dearth
of long-term monitoring and other data mean that the physical
mechanism responsible for this spectacular behaviour is still highly
unconstrained.
\end{abstract}


\section{Introduction}
\label{sect:intro}

Very large amplitude variations in the X-ray luminosity
(greater than a factor $\gs 100$) emanating from galactic nuclei are
unquestionably an indicator of unusual and interesting
phenomena. Such `transient-like' behaviour has been observed in only a
handful of 
galaxies to date (see Donley \et 2002 and Komossa 2002)
through observations with \rosat\footnote{Piro  \et 1988
reported a factor of $\gs 20$ decrease in the  $0.5-4.5$~keV
luminosity of E1615+061  between \heao\ and \einstein\  observations;
however, \asca\ measured rapid, persistent variability  in this object
(Guainazzi \et 1998) suggesting that it does not fit the above
definition.}.  
These soft X-ray bright galactic nuclei detected during the \rosat\ all-sky
survey (RASS) were found in subsequent follow-up observations with
the same satellite to be fainter, by factors of $70
- 400$, than their initial RASS detection.  Such spectacular long-term
fading is markedly different from the persistent variability usually
exhibited by soft X-ray bright Active Galactic Nuclei (AGN).

One possible explanation is that these sources are intrinsically
quiescent objects (dormant or inactive galactic nuclei) that were
subject to a luminous outburst near the  time of RASS, followed by a
decline in luminosity over timescales of months or years.
This scenario is explained by the tidal disruption model in which
the X-ray outburst is the result of a supermassive black hole in the
nucleus of a inactive galaxy capturing and  accreting a passing star
(e.g. Gurzadian \& Ozernoi 1980; Rees 1988).
Alternatively, a very different type of model is that in
which an otherwise normal (i.e. X-ray luminous) AGN suddenly
`switches  off,' possibly as the result of the thermal-viscous
accretion  disk instabilities known to operate in Galactic accreting
sources (Siemiginowska \et 1996). See Komossa (2002) and references
therein for further discussion of these and other models.

This letter presents exploratory \chandra\ observations of five of the
best-studied X-ray transient galactic nuclei.  The sample comprises
\wpvs, \ic, \rj, \ngc\ and \rxj.  Very little is known about the
post-outburst X-ray properties of these sources and these new
observations almost double the total number of post-RASS detections
for this sample.  The rest of this letter is organised as follows.
Section~\ref{sect:obs} describes the \chandra\ observations and X-ray
data analysis.  Section~\ref{sect:comp} discusses the optical source
classification and the construction of X-ray light curves by combining
these new \chandra\ data with earlier measurements.  Finally,
section~\ref{sect:disco} discusses the implications of these results.
Throughout this letter the cosmological parameters are assumed to be $
H_0 = 70 $~\kmpspMpc and $ q_0 = 0.5 $.

\section{Observations and Data Analysis}
\label{sect:obs}

\begin{figure*}
\hbox{
\includegraphics[width=5.0 cm, angle=270]{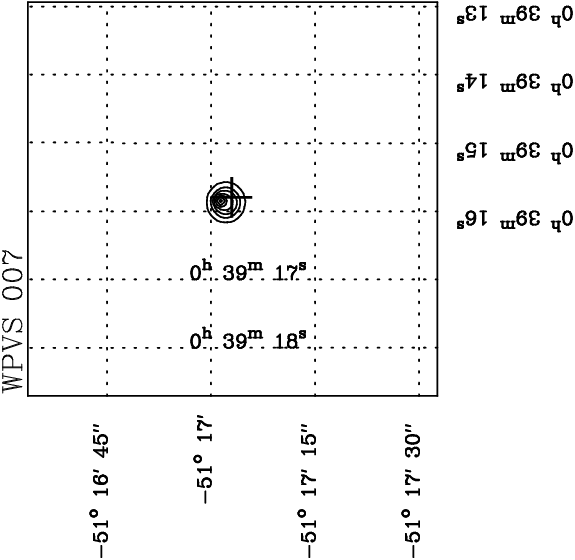}
\hspace{0.3 cm}
\includegraphics[width=5.0 cm, angle=270]{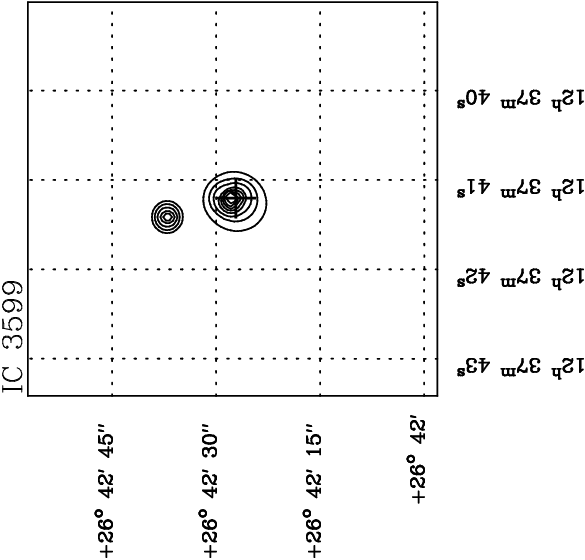}
\hspace{0.3 cm}
\includegraphics[width=5.0 cm, angle=270]{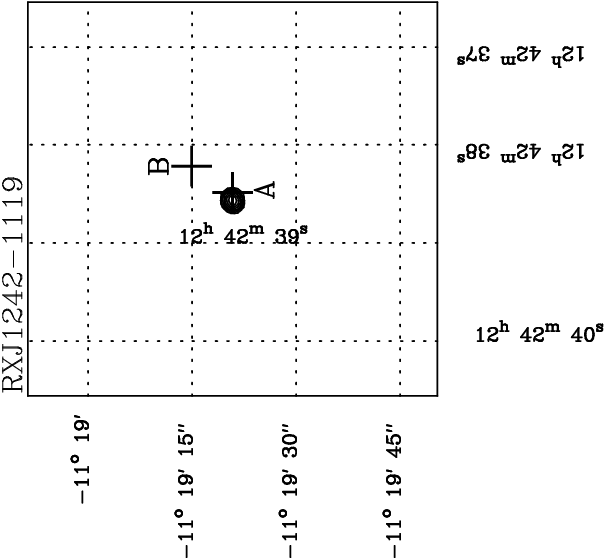}
}
\vspace{0.4 cm}
\hbox{
\hspace{3.9 cm}
\includegraphics[width=5.0 cm, angle=270]{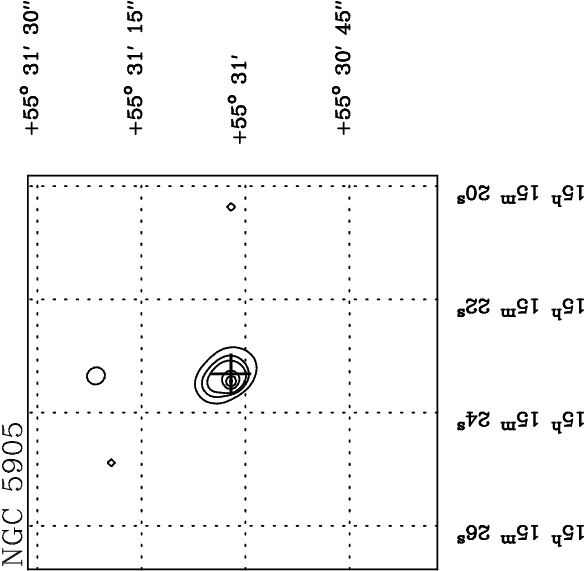}
\hspace{0.3 cm}
\includegraphics[width=5.0 cm, angle=270]{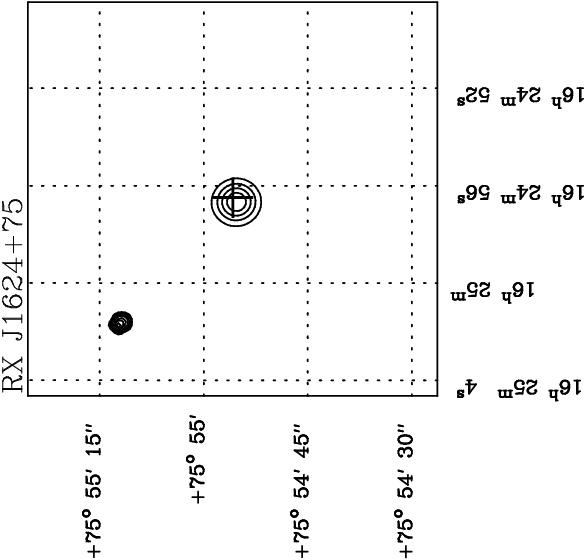}
}
\caption{
Contour plots of the X-ray intensity derived from
full-band ($0.3 - 7.0$~keV) ACIS-S3 images of the five target sources.
Each image spans $60$\arcsec$\times 60$\arcsec\ and has been adaptively
smoothed at the $2\sigma$ level (Ebeling, White \& Rangarajan
2003). Crosses mark the optical positions of the galactic nuclei. The
nucleus of each member of the galaxy pair is indicated in the case of
\rj.
}
\label{fig:images}
\end{figure*}

\subsection{Observations}

The five targets were each observed close to the aim-point of the
back-illuminated ACIS chip S3 (ACIS-S3).  As the expected X-ray fluxes
for these objects were uncertain by an order of magnitude, four of the
observations were performed with the ACIS CCDs using the quarter-frame
subarray. This reduced the CCD readout time (to $1.07$~s) and thereby
reduced any possible effects from photon pile-up (Ballet 1999) if the
sources were brighter than expected. The observation of \rj\ was
performed in full-frame mode.

The data were processed from the level-1 events files using {\tt CIAO
v2.3}.  Only events corresponding to grades 0, 2, 3, 4 and 6 were used
in the analysis of processed data. Flares in the background level were
identified by examining the light curve from the whole ACIS-S1 chip. These
showed that the observations of \wpvs, \ic\ and \rj\ were free from
background flares. The observation of \rxj\ showed an increase in the
background level in the final $\sim 200$~s of the observation; data
taken during this time interval were removed prior to analysis. The
observation of \ngc\ suffered from a higher background level (compared
to the other observations) such that the removal of the periods
of high background would leave insufficient data for
analysis. Therefore, the full exposure of \ngc\ was analysed accepting
that the background level was enhanced.  Table~\ref{tab:sources} lists
the basic properties of the five target sources and their \chandra\
observations.

\begin{table*}
\footnotesize
  \centering
  \caption{
Source observation log and properties.
}
\begin{tabular}{@{}lrrrrrrrrrr@{}}
\hline
Source          &R.A.           &Dec            &     &$N_{\rm H}~^{\rm a}$           &Observation&Exposure&      &  &  &  \\
Name            &(J2000)        &(J2000)        & $z$ &($10^{20}$~cm$^{-2}$) &Date       &(ks)  &Counts$^{\rm b}$  &SR$^{\rm c}$&Flux$^{\rm d}$&
$L_{\rm X}~^{\rm e}$ \\
\hline
\wpvs\          &00 39 15.8     &-51 17 03      & 0.029 & 2.6 &2002 Aug 2 &9.3     &$9.8\pm4.2$     & $2.4\pm1.7$       & $0.89$  & $2\times10^{40}$ \\
\ic\            &12 37 41.2     &26 42 29       & 0.022 & 1.3 &2002 Mar 7 &10.2    &$247.8\pm16.8$  & $5.2\pm0.9$       & $17$    & $2\times10^{41}$ \\
\rj$^{\rm f}$   &12 42 38.5     &-11 19 21      & 0.05  & 3.6 &2001 Mar 9 & 4.5    &$17.9 \pm5.3$   & $1.3\pm0.6$       & $2.8$   & $2\times10^{41}$ \\
\ngc\           &15 15 23.4     &55 30 57       & 0.011 & 1.4 &2002 Oct 4 &9.6     &$25.3\pm6.6$    &  $1.3\pm0.6$      & $1.8$   & $6\times10^{39}$ \\
\rxj\           &16 24 56.5     &75 54 56       & 0.064 & 3.8 &2002 Sep 15 &10.1   &$3.5\pm3.2$     &   $0.3\pm0.4$     & $0.24$  & $3\times10^{40}$ \\
\hline
  \end{tabular}

\raggedright
$^{\rm a}$ Galactic column density from Dickey \& Lockman (1990).
$^{\rm b}$ Larger of the two $1\sigma$ error bounds using the approximation of Gehrels (1986).
$^{\rm c}$ SR is the $0.3-1.0/1-7$~keV softness ratio.
$^{\rm d}$ Flux in the $0.3-7.0$~keV band ($10^{-14}$ erg s$^{-1}$ cm$^{-2}$).
$^{\rm e}$ The estimated unabsorbed X-ay luminosity in the same band (erg s$^{-1}$).
$^{\rm f}$ Optical position of galaxy A (Komossa \& Greiner 1999).
\label{tab:sources}
\end{table*}

\subsection{X-ray Imaging}

Figure~\ref{fig:images} shows the X-ray contour plots derived from the
ACIS images for the five targets
after adaptive smoothing has been applied. Clearly in all cases
there is an excess of photons coincident with the optical position of
the galactic nucleus (i.e. within the expected $\sim 2$\arcsec\
uncertainty in the optical position).  The optical positions were
taken from the references given in Section~\ref{sect:comp} and the
NASA/IPAC Extragalactic Database\footnote{{\tt http://nedwww.ipac.caltech.edu/}} 
(NED).

In the case of \rj\ the positional error circle of the RASS X-ray
source  contains a pair of inactive galaxies, labelled A and B by
Komossa \& Greiner (1999). From the \rosat\ data it was not clear which
of the two galaxies should be identified with the X-ray source. The
X-ray source detected by the \chandra\ observation is clearly
coincident with galaxy A (Fig.~\ref{fig:images}). Assuming that this
is the same X-ray source as detected by \rosat\ then the X-ray
outburst should be associated with galaxy A. All the X-ray sources
appear consistent with being point-like with the possible exception of
\ngc, which displays a slightly asymmetric shape in the smoothed
image.  However, with only $\sim25$ counts in the source and an
enhanced background level it is difficult to be more confident of
this without a more sensitive observation. It is interesting to
note that of the five targets \ngc\ is the nearest ($1$\arcsec or $2$
pixels corresponds to a spatial scale of $\sim 0.2$~kpc at the
source redshift) and thus might be most likely to show extended
emission in the \chandra\ images.

\subsection{Count Rates and Softness Ratios}
\label{sect:cr}

Source counts were estimated by performing photometry on the raw
(unsmoothed) images with a circular aperture of radius of $2$\arcsec
centred on the source. The background level was estimated from a
concentric annulus with inner and outer radii of $3$\arcsec\ and
$60$\arcsec, respectively (excluding nearby sources where present).
The net counts associated with each source are given in
Table~\ref{tab:sources}.  The brightest object, \ic, has $\gs 200$
photons in its X-ray image, enough for crude spectral analysis (see
below). The other four detections are based on far fewer counts. In the case
of \rxj\ there are only $4$ counts in the source aperture, i.e. a
$\sim 2\sigma$ confidence detection (with negligible background), but
the coincidence with the optical position makes this a likely
detection of the nuclear X-ray emission.

Softness ratios were calculated for use as a crude indicator of the
spectral slopes of the faint sources. The softness ratio was defined
as the ratio of counts in the bands $0.3-1.0/1.0-7.0$~keV. These are
given in Table~\ref{tab:sources} and show \ic\ to have a very soft
spectrum with
the other four sources showing somewhat harder emission.  
For the three harder sources (\rj, \ngc\ and \rxj) the
softness ratio corresponds to a power-law photon index $\Gamma \le
2.5$.  The measured softness ratio of \ic\ requires $\Gamma \sim 4$
(see below). 

In all cases the
$0.3 - 7.0$~keV flux was estimated using the {\tt PIMMS} calculator at
the \chandra\ X-ray Center\footnote{{\tt
http://asc.harvard.edu/}}. For this purpose the spectrum was assumed
to be a power-law 
(with a photon index $\Gamma=3$) modified by Galactic absorption. The
estimated fluxes are listed in Table~\ref{tab:sources}. 
If the underlying X-ray spectra differ substantially from the assumed
spectral form then these flux estimates may change by a factor of
a few.
The $0.3-7.0$~keV unabsorbed luminosities were
estimated assuming the same spectral model and are also listed in the
table.

\subsection{X-ray Spectrum of \ic}

Spectra were extracted from the source and background regions of the
observation of \ic. A response matrix and an ancillary response file
were generated with {\tt mkrmf} and {\tt mkwarf}, respectively. The
low number of counts meant that applying standard binning
(i.e. $N\ge20$ counts per energy bin) would result in too few bins for
spectral fitting. Therefore the unbinned spectrum was fitted by
minimising the $C$-statistic (Cash 1979), appropriate for situations
when spectra contain few counts. The fitting was performed in {\tt
XSPEC v11.2} (Arnaud 1996). 

\begin{figure}
\includegraphics[width=5.80 cm, angle=270]{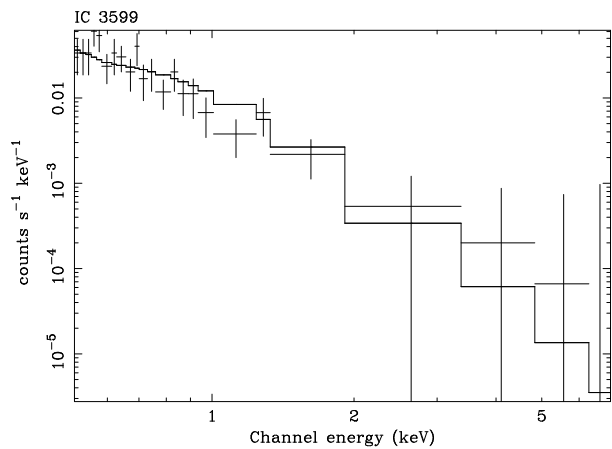}
\caption{
ACIS-S3 spectrum of \ic\ (crosses) fitted with a power-law model
(histogram).
The data have been rebinned for display purposes.
}
\label{fig:spectrum}
\end{figure}

The spectral fitting was restricted to the $0.6 - 7.0$~keV band since
the ACIS calibration is uncertain below $0.6$~keV.  A power-law model
with Galactic absorption gave a best-fitting photon index of $\Gamma =
3.56_{-0.34}^{+0.37}$ ($90$ per cent confidence limits) which is 
steeper than that
normally seen in Seyfert galaxies. This fit is shown in
Figure~\ref{fig:spectrum}. Fitting with alternative spectral models
(blackbody or bremsstrahlung continuum or a {\tt mekal} plasma model)
gave noticeably larger data/model residuals. The best-fitting temperatures
were $kT \sim 0.16$~keV for a blackbody and $\sim 0.26$~keV for a 
{\tt mekal} plasma model.

\section{Comparison with other observations}
\label{sect:comp}

\subsection{X-Ray light curves and softness ratios}

\begin{figure}
\centering
\includegraphics[width=7.5 cm, angle=0]{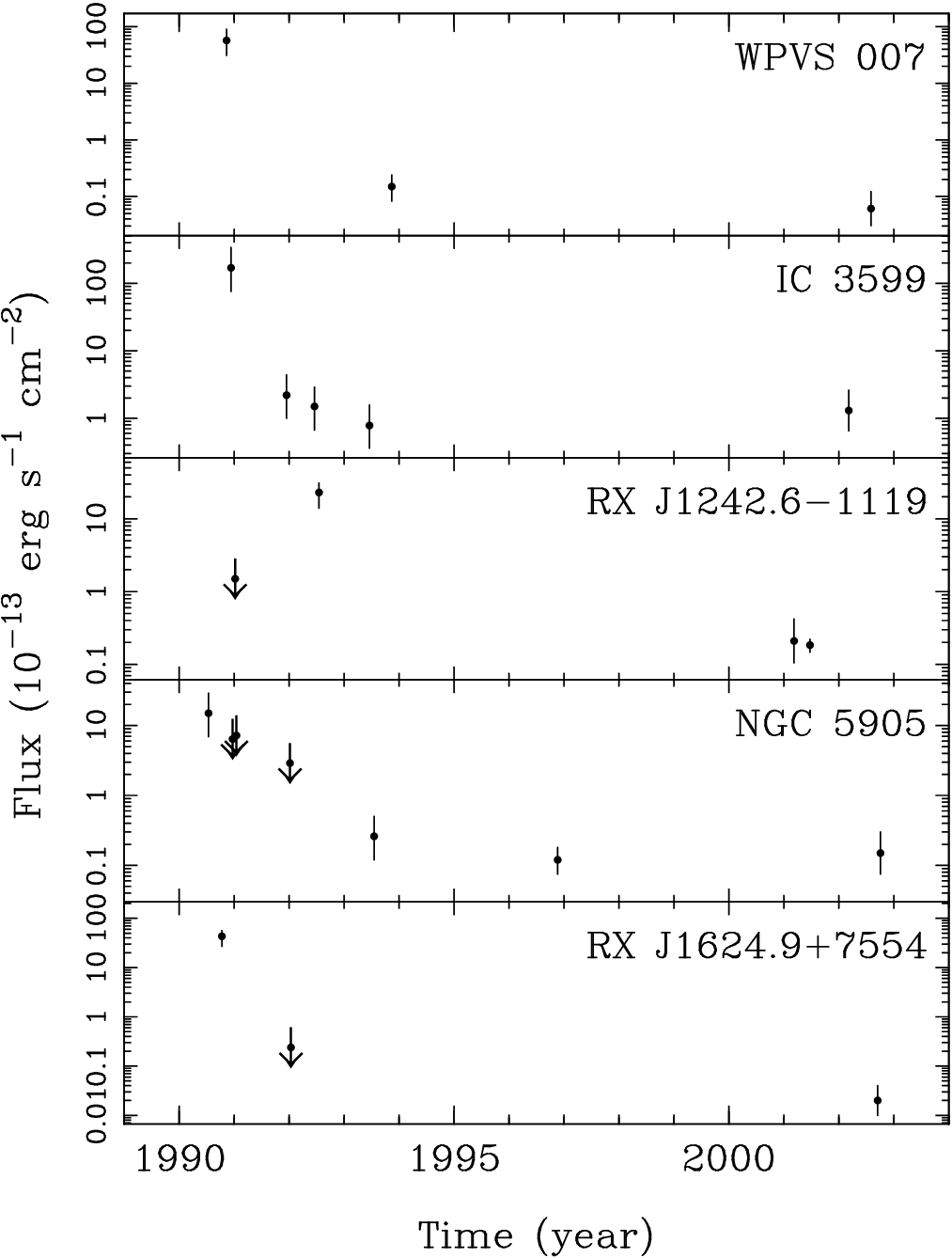}
\caption{Historical light curves of soft X-ray ($0.3 - 2.0$~keV) flux
compiled from \rosat\ PSPC observations (left hand side
flux points) and
new \chandra\ observations (data points on far right). 
In addition, the fluxes derived from one \rosat\ HRI observation of
NGC 5905 (1996.9) and one \xmm\ observation of \rj\ (2001.5) are also
shown. Upper limits
on the flux, derived from non-detections with \rosat\ are marked
with arrows.}
\label{fig:lc}
\end{figure}

In order to better understand how these sources have changed since
their X-ray outbursts, long term light curves were constructed for
each of the five sources, as shown in Fig.~\ref{fig:lc}.   For this
purpose the  \chandra\ $0.3-7.0$~keV count rates were converted into
$0.3-2.0$~keV fluxes using {\tt PIMMS}, assuming the energy spectrum
is a steep power-law (with $\Gamma=3$) modified by Galactic
absorption.  The  $0.3-2.0$~keV band was chosen as both the \chandra\
ACIS and \rosat\ PSPC cover this energy range with reasonable
effective area.   The \rosat\ data points were derived from both the
RASS and  pointed observations of each source (see below for
references).  The $0.1-2.4$~keV PSPC count rates were converted to
$0.3-2.0$~keV fluxes assuming the same spectrum as above.  In addition
a single \rosat\ HRI observation of \ngc, taken in 1996 October, is
included.  Since the largest uncertainty associated with these flux
estimates is due to the model-dependence of the counts-to-flux
conversion the fluxes were also calculated assuming $\Gamma=2$ and
$\Gamma=4$. This resulted in flux estimates differing by factors $\gs
2$ from the fiducial model and the error bars shown in
Fig.~\ref{fig:lc} reflect this large uncertainty.

\xmm\ observed \rj\ during 2001 June 21-22, for a duration of
$30.1$~ks, and detected the X-ray source (with $\approx 286$ counts). A
spectrum was extracted from the EPIC pn data (using standard
procedures) and fitted in {\tt XSPEC}. This was well-fitted
with a simple absorbed power-law model ($\Gamma = 2.42 \pm 0.23$, consistent
with the estimates from section~\ref{sect:cr}). The $0.3-2.0$~keV X-ray
flux was well-constrained at $\approx 1.8 \times 10^{-14}$~erg
s$^{-1}$ cm$^{-2}$, consistent with the \chandra\ measurement.
This flux point is also shown in Fig.~\ref{fig:lc}.

As can be seen, for all five sources the \chandra\ fluxes lie at least
two orders of magnitude below the brightest measured \rosat\
fluxes. None have shown a significant increase in flux at any time
during the (albeit poorly-sampled) period since the huge decline in
flux was first established. These results are all consistent with a
decline from high luminosity to a persistent low luminosity ($L_{\rm
X} \sim 10^{40}-10^{41}$~erg s$^{-1}$) over a timescale $\ls 2$ years.

The \chandra\ softness ratios also indicate changes.  For all five
objects the \rosat\ X-ray spectrum was extremely soft during outburst.
In fact \wpvs\ was the softest spectrum AGN observed during the RASS
(Grupe \et 1995b) with an effective photon index $\Gamma \sim 8$. The
other objects showed slopes in the range $\Gamma = 3 - 5$ (Brandt,
Pounds \& Fink 1995; Komossa \& Greiner 1999; Grupe, Thomas \& Leighly
1999; Bade, Komossa \& Dahlem 1996). In the case of \ic\ the spectrum
remains rather soft.  For the other four objects the \chandra\
softness ratios  suggest harder spectra ($\Gamma \ls 2.5$). However,
it is difficult to make a more quantitative comparison due to the
different  energy ranges of the \rosat\ and \chandra/ACIS spectra and
the small number of source photons in the \chandra\ data.

\subsection{Optical Spectra and Classification}

\noindent
{\bf \wpvs: }
Two years prior to its RASS detection an optical spectrum 
was obtained by Winkler,  Stirpe \& Sekiguchi (1992). They classified
the source ($\#7$ in their notation) as a Seyfert 1.   
Observations  by Grupe \et (1995b) taken two years after the RASS
detection identified  \wpvs\ as a  Narrow-line Seyfert 1 galaxy (NLS1;
Boller, Brandt \& Fink 1996). An \hst/FOS ultraviolet spectrum taken
in 1996 also showed broad permitted lines as well as intrinsic, ionised
absorption (Crenshaw \et 1999; Goodrich 2000).

\noindent
{\bf \ic: }
The optical spectrum taken five months after its initial RASS
detection showed \ic\ to be a NLS1 (Brandt \et 1995). However, 
further observations 14 months after the RASS detection showed
the optical spectrum to have changed to resemble that of a Seyfert 1.9
(Grupe \et 1995a).  Observations in subsequent years confirmed this
(Grupe \et 1995a; Komossa \& Bade 1999).

\noindent
{\bf \ngc: } This source was
identified as an \hii/starburst galaxy six years after its RASS
detection (Bade, Komossa \& Dahlem 1996; Komossa \& Bade 1999).  
However, higher spatial resolution
spectroscopy with \hst\ revealed narrow, high-ionisation lines
originating from the nucleus (Gezari \et 2003).  These imply the
presence of a low-luminosity Seyfert 2 nucleus that was swamped by the
surrounding \hii\ emission in the previous observations.  Gezari \et
(2003) used the correlation between H$\alpha$ and soft X-ray luminosity
described by Halderson \et (2001) to estimate the soft X-ray
luminosity of the nucleus to be $L_{0.1-2.4} \sim 4 \times
10^{40}$~erg s$^{-1}$, a factor $\ls 6$ larger than the X-ray
luminosity actually observed by \chandra.

\noindent
{\bf  \rj\ and \rxj: } 
These appeared as otherwise inactive galaxies when observed several
years after their initial \rosat\ detections (Komossa \& Greiner 1999;
Grupe, Thomas \& Leighly 1999).
Gezari \et (2003) report no detectable non-stellar
continuum or high-ionisation line emission in their \hst\ observations.

\section{Discussion}
\label{sect:disco}

As is clear from the light curves, in all five cases the flux recorded
by \chandra\ falls $2-3$ orders of magnitude below the maximum flux
observed by \rosat. Although the light curves are sparsely sampled
this does strongly suggest that the X-ray light curves are best
characterised by a single, dramatic decline in luminosity 
on a timescale of $\ls 2$ years followed by a period of relative
quiescence. This is consistent with the X-ray outbursts
being non-recurring events (at least on timescales $\ls 10$ years).
In the most extreme example, \rxj\ faded by a factor $\gs 1000$
between its RASS observation in 1990 October and its \chandra\
observation in 2002 September. The optical spectra seem to indicate a
wide variety of source types, ranging from genuine AGN to inactive galaxies.

It is difficult to make a simple comparison with the fading predicted in
the tidal disruption scenario (Rees 1988): $L_{\rm X} \propto
(t-t_{0})^{-5/3}$ (where $t_{\rm 0}$ is the time of the outburst
event).  The \chandra\ X-ray luminosities could contain a significant
contribution from unrelated galactic emission such as star forming regions,
bright X-ray binaries, diffuse emission components, etc. 
Indeed, individual Ultra-Luminous X-ray (ULX) sources in nearby
galaxies can reach X-ray luminosities $\sim 10^{40}$~erg s$^{-1}$
(Fabbiano \& White 2003).
This `background' emission is an unknown quantity and, if dominated by a few bright
X-ray binaries, could also be variable.  Thus the X-ray light curve
following stellar disruption should be: $L_{\rm X} \sim N (t -
t_{x})^{-5/3} + C$ where $C$ is the unknown background galactic
emission. The model therefore has three unknowns ($N$, $t_{\rm x}$ and
$C$) but the light curves unfortunately have only $2-5$ data points
making the test rather meaningless. In the best sampled light
curves (\ic\ and \ngc) the last \rosat\ flux is comparable to the
\chandra\ flux, implying no further fading has occurred on a timescale
of $\sim 10$~years.  Unfortunately, in all five cases, 
it is not clear whether the quiescent source is a residual 
low-luminosity nuclear source or unrelated, background galactic emission.

The question of whether the `switch off' marked the end of a
single, isolated accretion episode (such as a tidal disruption event)
or a rapid decrease in the luminosity of a persistent AGN (Seyfert
galaxy to LLAGN) remains largely open. 
The peak luminosities were $L_{\rm X} \gs 10^{43}$ erg s$^{-1}$,
comparable with bright Seyfert 1 galaxies, while the quiescent
luminosities are only $L_{\rm X} \sim 10^{40} - 10^{41}$~erg s$^{-1}$.
The latter are rather
high compared to the nuclear emission expected from normal/inactive
or starburst galaxies (see Fabbiano 1989) but quite comparable to those of
low-luminosity AGN (LLAGN: Ptak \et 1999; Roberts \& Warwick 2000; Ho \et 2001; Ptak 2001).  
Thus the evidence does favour the presence of LLAGN in these galaxies.
The existence of long-lasting, high-ionisation optical line emission
from the nuclei of \wpvs, \ic\ and \ngc\ further suggests these may harbour
some kind of long-lived AGN. However, it is difficult to make any strong claims
about their prior X-ray activity since no suitable observations exist.

The remaining two galaxies, \rj\ and \rxj, show no evidence for a
luminous AGN in their optical spectra and remain the best candidates for
tidal disruption 
events, athough their (relatively) high residual X-ray luminosity may 
indicate some residual nuclear activity (as argued above).  The
optically inactive galaxy \rj\ is particularly interesting as this was
the only one of the target sources to have been observed  (although
not detected) just prior to its outburst detection (see Komossa \&
Greiner 1999). This suggests a rise time for the outburst of less than two years.

Transient galactic nuclei represent a relatively new and exciting
avenue of X-ray astronomy research (see Komossa 2002 for a review). So
little is known about these objects that any future X-ray observations
are  potentially of great importance. For example, a longer
observation of \ngc\ could reveal whether the X-ray emission is
extended (and hence due to the circumnuclear starburst, not a
LLAGN). Deeper observations of \ic\ would better define the X-ray
spectrum and thereby help clarify the origin of the remaining
low-luminosity  X-ray emission. Future monitoring of these sources is
needed to see whether they are persistently variable (which would
imply on-going accretion) and, in particular, to see whether any show
repeat outbursts.  A severe hindrance to such a project  is the dearth
of known sources. Future large-area monitoring missions such as
\lobster\ (Fraser 2002) are well suited to finding transient galactic
nuclei and providing the most likely route to a reasonably sized
sample of such objects (see discussions in Sembay \& West 1993 and
Donley \et 2002) on which a concerted programme of follow-up
observations might be based.

\section*{Acknowledgements}
We thank Steve Allen for help with {\tt CIAO}, Tim Roberts for
useful discussions and an anonymous referee for a helpful
report. SV acknowledges financial support from PPARC.
This research has made use of the NASA/IPAC Extragalactic Database
(NED) which is operated by the Jet Propulsion Laboratory, California
Institute of Technology, under contract with the National Aeronautics
and Space Administration.

\bsp
\label{lastpage}
\end{document}